\documentclass[12pt,preprint]{aastex}

\newcommand{\kepler}{{\it Kepler}}

\shorttitle{\kepler{} Mission Signal Detections}
\shortauthors{Tenenbaum et al.}

\slugcomment{Draft: \today}
\begin{document}

\title{Detection of Potential Transit Signals in Sixteen Quarters
of \kepler{} Mission Data}

\author{Peter Tenenbaum, Jon M. Jenkins, Shawn Seader, Christopher J. Burke, 
Jessie L. Christiansen\footnote{Current affiliation: NASA Exoplanet Science Institute, 
California Institute of Technology, Pasadena, CA 91125, USA}, 
Jason F. Rowe, 
Douglas A. Caldwell, Bruce D. Clarke, 
Jeffrey L. Coughlin, 
Jie Li, Elisa V. Quintana, 
Jeffrey C. Smith, 
Susan E. Thompson, and Joseph D. Twicken}
\affil{SETI Institute/NASA Ames Research Center, Moffett Field, CA 94305, USA}
\email{peter.tenenbaum@nasa.gov}
\author{ 
Michael R. Haas, Christopher E. Henze, Roger C. Hunter, and Dwight T. Sanderfer}
\affil{NASA Ames Research Center, Moffett Field, CA 94305, USA}
\author{Jennifer R. Campbell, Forrest R. Girouard, 
Todd C. Klaus, 
Sean D. McCauliff, 
Christopher K. Middour, 
Anima Sabale, Akm Kamal Uddin, and Bill Wohler}
\affil{Orbital Sciences Corporation/NASA Ames Research Center, Moffett Field, CA 94305, USA}
\and
\author{Thomas Barclay and Martin Still}
\affil{BAER Institute/NASA Ames Research Center, Moffett Field, CA 94305, USA}

\begin{abstract}
We present the results of a search for potential transit signals in four years of photometry data
acquired by the \kepler{} Mission.  The targets of the search include 111,800 stars which were
observed for the entire interval and 85,522 stars which were observed for a subset of the
interval.  We found that 9,743 targets contained at least one signal consistent with the
signature of a transiting or eclipsing object, where the criteria for detection are periodicity of the
detected transits, adequate signal-to-noise ratio, and acceptance by a number of tests which
reject false positive detections.  When targets that had produced a signal
were searched repeatedly, an additional 6,542 signals were
detected on 3,223 target stars, for a total of 16,285 potential detections.
Comparison of the set of detected signals with a set of known and vetted transit events
in the \kepler{} field of view shows that the recovery rate for these signals is 96.9\%.  The ensemble
properties of the detected signals are reviewed.
\end{abstract}



\section{Introduction}

We have reported on the results of past searches of the \kepler{} Mission data for signals of transiting
planets, in particular the search of the first three quarters \citep{pt2012} and the search of the first three
years \citep{pt2013}.  We now update and extend those results to incorporate an additional year of 
data acquisition and an additional year of \kepler{} Pipeline development. 

\subsection{\kepler{} Science Data}

The details of \kepler{} operation and data acquisition have been reported elsewhere \citep{science-ops}.
In brief:  the \kepler{} spacecraft is in an Earth-trailing heliocentric orbit and maintained a boresight pointing centered 
on $\alpha = 19^{\rm h}22^{\rm m}40^{\rm s}, 
\delta = +44.5\degr$.  The \kepler{} photometer acquired data on a 115 square degree region of the sky.  The data
were acquired in 29.4 minute integrations, colloquially known as ``long cadence'' data.  The spacecraft was required
to rotate about its boresight axis by 90 degrees every 93 days in order to keep its solar panels and thermal
radiator correctly oriented; the interval which corresponds to a particular rotation state is known colloquially 
as a ``quarter.'' Because of the quarterly rotation, target stars were observed throughout the year in 4 different
locations on the focal plane. Science acquisition was interrupted monthly for data downlink, quarterly for
maneuver to a new roll orientation (typically this is combined with a monthly downlink to limit the loss of 
observation time), once every 3 days for reaction wheel desaturation (one long cadence sample is sacrificed
at each desaturation), and at irregular intervals due to spacecraft anomalies. In addition to these interruptions
which were required for normal operation, data acquisition was suspended for 11.3 days, from 2013-01-17 19:39Z
through 2013-01-29 03:50Z (555 long cadence samples)
\footnote{Time and date are presented here in ISO-8601 format, YYYY-MM-DD HH:MM, or optionally 
YYYY-MM-DD HH:MM:SS, with a trailing `Z' to denote UTC.}
: during this time, the spacecraft reaction wheels
were commanded to halt motion in an effort to mitigate damage which was being observed on Reaction Wheel
4, and spacecraft operation without use of reaction wheels is not compatible with high-precision photometric
data acquisition.

In July 2012, one of the four reaction wheels used to maintain spacecraft pointing during science acquisition
experienced a catastrophic failure. The mission was able to continue, using the remaining three wheels to 
permit 3-axis control of the spacecraft, until May of 2013. At that time a second reaction wheel failed, which 
forced an end to \kepler{} data acquisition in the nominal \kepler{} field of view. As a result, the analysis reported
here is the first which incorporates the full volume of data acquired from that field of view.

\kepler{} science data acquisition began at 2009-05-12 00:00Z, and acquisition of Quarter 16 data
concluded at 2013-04-08 11:17:11Z. This time interval contains 69,810 long cadence intervals.  Of these,
4,726 were consumed by the interruptions listed above, and 65,084 long cadence intervals were dedicated to
science data acquisition.  An additional 1,076 long cadence intervals were excluded from use 
in searches for transiting planets. These
samples were excluded due to data anomalies which came to light during processing and inspection of the 
flight data.  This includes a contiguous set of 255 long cadence samples acquired over the 5.2 days which immediately 
preceded the 11 day downtime described above: the shortness of this dataset combined with the duration of the
subsequent gap led to a judgement that the data would not be useful for transiting planet searches.

\begin{figure}
\plotone{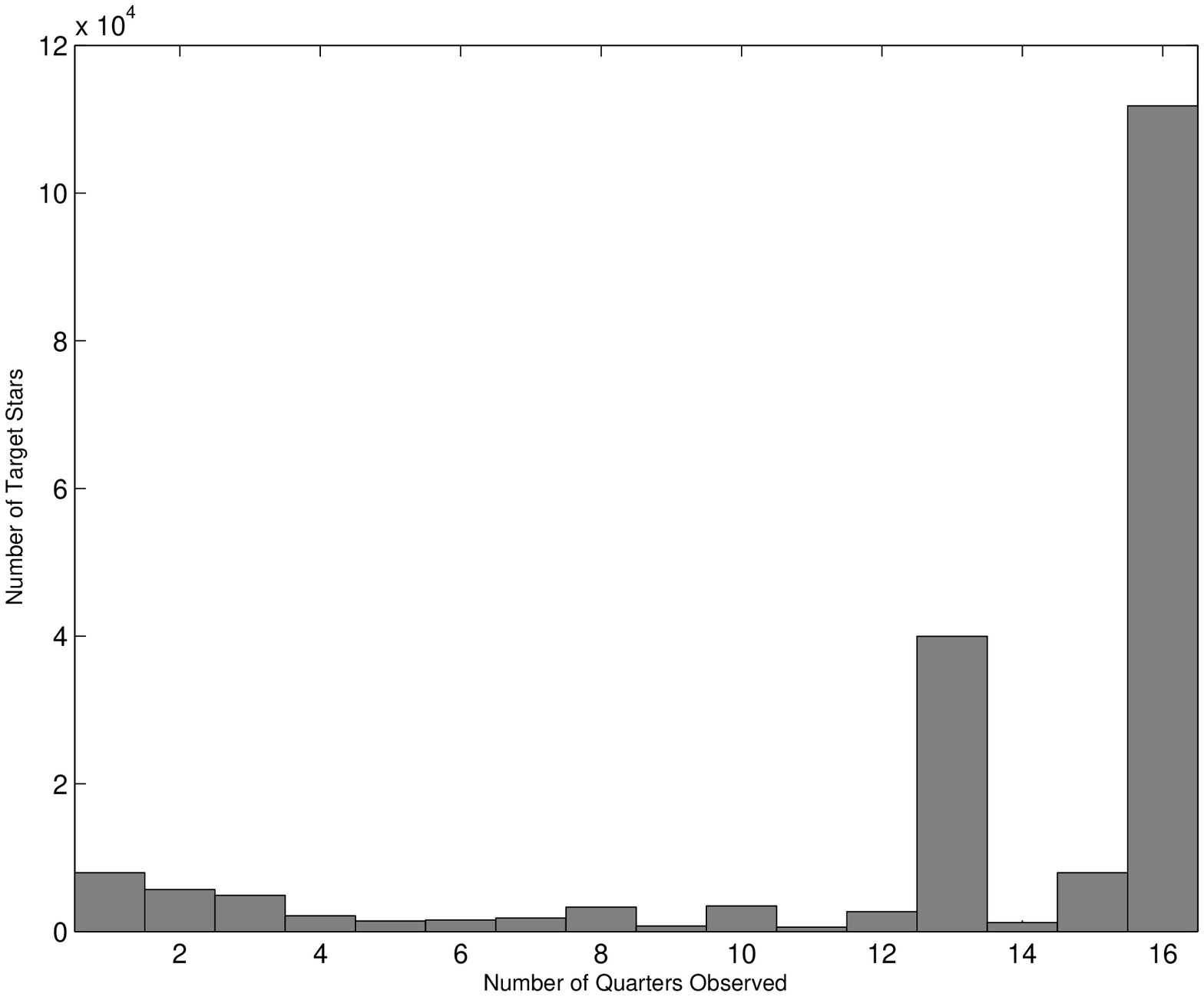}
\caption{Histogram of number of quarters of observation for all targets.
\label{f1}}
\end{figure}

A total of 197,322 targets observed by \kepler{} were searched for evidence of transiting planets. This set of targets
includes all stellar targets observed by \kepler{} at any point during the mission, and specifically includes target
stars which were not originally observed for purposes of transiting planet searches (asteroseismology targets, guest
observer targets, etc.). The exception to this is a subset of known eclipsing binaries, as described below.
Figure \ref{f1} shows the distribution of targets according
to the number of quarters of observation. A total of 111,800 targets were observed for all 16 quarters. An additional
39,964 targets were observed for 13 quarters: the vast majority of these targets were in regions of the sky which
are observed in some quarters by CCD Module 3, which experienced a hardware failure in its readout electronics
during Quarter 4, resulting in a ``blind spot'' which rotates along with the \kepler{} spacecraft.  The balance of 45,558
targets which were observed for some other number of quarters is largely due to gradual changes in the target
selection process over the duration of the mission.

As described in \citet{pt2013}, some known eclipsing binaries are excluded from planet searches. In this case, a total of
1,519 known eclipsing binaries were excluded. This number is smaller than the number excluded in the Q1-Q12
analysis due to a change in exclusion criteria. Specifically, we excluded eclipsing binaries from the most recent 
Kepler catalog of eclipsing binaries \citep{eb-cat} that did not meet the criteria of being ``transit-like''. 
Eclipsing binaries were considered to be transit-like only if all of the following criteria were met:

\begin{enumerate}
\item The primary eclipse depth is greater than or equal to zero, i.e., flux must decrease at primary eclipse
\item The primary and secondary eclipse depths are within 10\% of each other OR the secondary eclipse depth is less than 10\% of the primary eclipse depth
\item If detected, the phase of secondary eclipse has to occur within the range 0.49 - 0.51, i.e., the binary star's orbit must be near-circular
\item The morphology of the system, as defined in \citet{eb-cat}, 
has to be $<$ 0.6, i.e., the primary and secondary eclipses must be well separated from one another.
\end{enumerate}

\noindent Thus, the excluded eclipsing binaries are largely
contact binaries which produce the most severe misbehavior in the Transiting Planet Search (TPS)
pipeline module \citet{pt2012,pt2013}, 
while well-detached, transit-like eclipsing binaries
are now processed in TPS.  This was done in order to ensure that no possible transit-like signature was excluded, 
and also to produce examples of the outcome of processing
such targets through both TPS and Data Validation \citet{hw2010,jt2014}, such that quantitative differences
between planet and eclipsing binary detections could be determined and exploited for rejecting other,
as-yet-unknown, eclipsing binaries detected by TPS.

\subsection{Processing Sequence: Pixels to TCEs to KOIs}

The steps in processing \kepler{} science data have not changed since \citet{pt2012}, and are briefly summarized below.
The pixel data from the spacecraft were first calibrated to remove pixel-level effects such as gain variations, linearity, 
and bias. The calibrated pixel values were then combined within each target to produce a flux time series for that target.
The ensemble of target flux time series were then corrected for systematic variations driven by effects such as 
differential velocity aberration, temperature changes, and small instrument pointing excursions. These corrected flux
time series became the inputs for the Transiting Planet Search.

The Transiting Planet Search (TPS) software module analyzed each corrected flux time series individually for evidence of
periodic reductions in flux which would indicate a possible transiting planet signature. The search process incorporated
a significance threshold (often referred to as the "multiple event statistic")
and a series of vetoes; the latter were necessary because while the significance threshold was sufficient
for rejection of the null hypothesis, it was incapable of discriminating between multiple competing alternate hypotheses
which can potentially explain the flux excursions. An ephemeris on a given target which satisfied the significance threshold
and passed all vetoes is known as a Threshold Crossing Event (TCE). Each target with a TCE was then searched for
additional TCEs, which potentially indicated multiple planets orbiting a single target star.

After the search for TCEs concluded, additional automated tests were performed to assist members of the Science Team 
in their efforts to reject false positives. A TCE which has been accepted as potential transiting planets, based on analysis of these
additional tests, is designated as Kepler Objects of Interest (KOIs). Note that, while the TCEs were determined in a purely algorithmic
fashion by the TPS software module, KOIs have been selected on the basis of examination and analysis by human scientists.

\subsection{Pre-Search Processing}

Since the publication of \citet{pt2013}, there have been considerable improvements to the
Pre-Search Data Conditioning (PDC) component of the Kepler pipeline. The purpose of PDC is to remove variations in the
flux time series which are generated by changes in the spacecraft environment or other systematic effects. PDC performs
very well for the majority of targets in the Kepler Field of View. However, for an appreciable minority 
the Bayesian Maximum A Posteriori (PDC-MAP) algorithm \citep{js2012} does not produce acceptable corrections of 
the visible systematics.
To further minimize the number of targets for which PDC-MAP fails to perform admirably, a new method has been developed:
multi-scale MAP (or msMAP) (M.C. Stumpe et al., 2014, in preparation).
Utilizing an overcomplete discrete wavelet transform the new method divides
each light curve into multiple channels, or bands, based upon characteristic signal scales in time and frequency.
This produces three time series for each flux time series: one dominated by each of short-timescale ($\le3$ long cadences), 
medium-timescale (4 to 1023 long cadences),
and long-timescale (1024 or more long cadences) variations.
The PDC-MAP algorithm is then applied to each band separately, which
allows for a better separation of characteristic signals and cleaner removal of the systematics. Relevant to transit
detection, the new msMAP provides two distinct improvements to the PDC processed data. The first is a significantly
improved removal of thermal transients which occur in the transition from Earth-pointing to science pointing after each data downlink. The
second is a modest reduction in introduced noise. 

A second significant improvement to PDC is that it now ``protects'' known transits from false detection as Sudden Pixel
Sensitivity Dropouts (SPSDs) or other types of outlier. Cadences containing known transits and eclipses are computed using the
known epoch, period and duration of the events, and assuming a linear ephemeris. No SPSDs or outliers are flagged during the
known transits. This helps preserve transit depths and shapes from corruption by the SPSD and outlier correction algorithms. 
Note that this only
affects known transits. There is still the risk of transit corruption for as yet undetected transits. However, once the
transits are detected and validated, subsequent data processing iterations will incorporate the new information.

\section{Transiting Planet Search}

This section describes the changes which have been made to the TPS algorithm since \citet{pt2013}.  For
further information on the algorithm, see \citet{jmj2002}, \citet{jmj2010}, and \citet{pt2012}.

\subsection{Removal of Positive Flux Outliers}\label{remove-positive-outliers}

As described in Section 2.4 of \citet{pt2013}, removal of negative flux outliers is a hazardous
action, since it relies upon an algorithmic capability to distinguish between a true outlier and a transit, 
and for obvious reasons removing the latter is frowned upon.  For this reason, strict limitations are
placed upon the algorithm's capabilities for removing suspected negative outliers.

\begin{figure}
\plotone{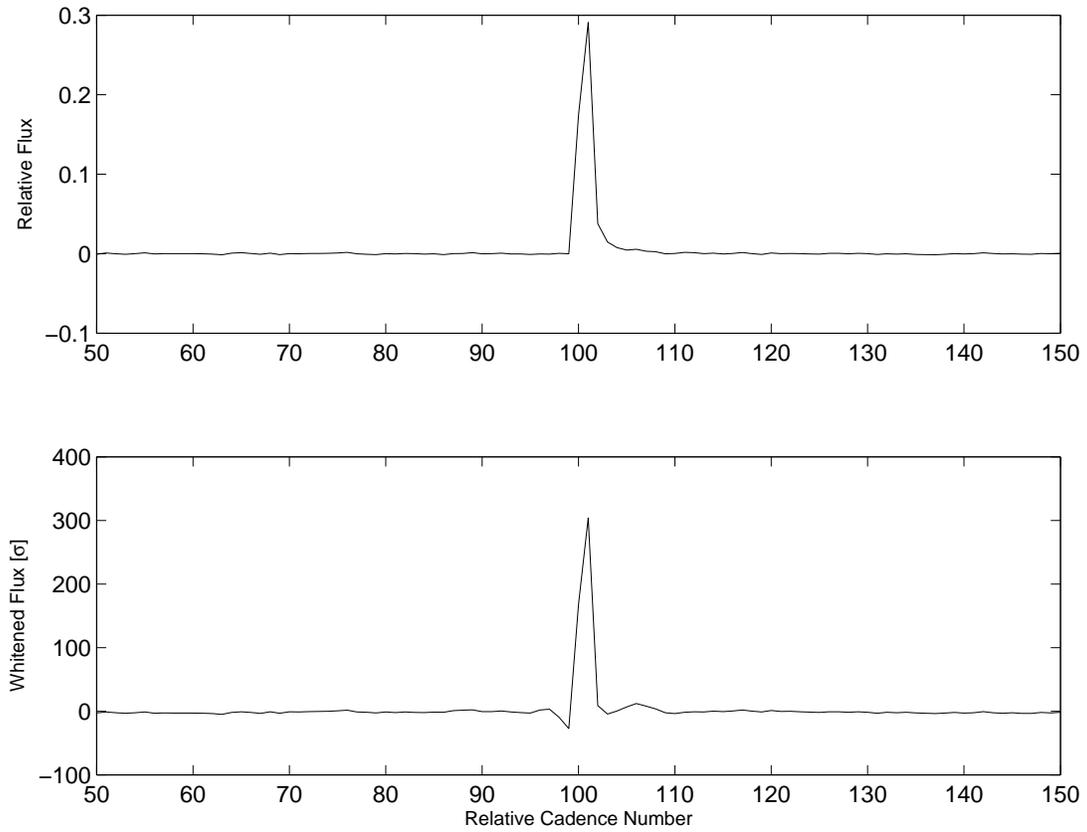}
\caption{Effect of a positive flux outlier.  Top: original flux.  Bottom: whitened flux. Note that whitening introduces a negative outlier
to the whitened flux, which can be misconstrued as a transit.  While the resulting negative outlier is much smaller than the original 
positive outlier, in this case the negative outlier still has a single event significance of over 27 $\sigma$.
\label{f2}}
\end{figure}

Positive outliers are much less risky to remove, since by definition a positive outlier looks like the opposite of
a transit.  At first glance, one might therefore assume that positive flux outliers are irrelevant as a source
of false alarms or other difficulties, since the difference between a short-duration positive flux
excursion and a short-duration negative flux excursion is intuitively obvious to the most casual observer.
In actuality, however, positive flux excursions can result in false alarm detections via the following mechanism:
when a positive flux excursion is subjected to the whitening filter, the whitened result includes ringing
which precedes and follows the excursion, as shown in Figure \ref{f2}. The strongest components of the
ring-down have the opposite sign to the original excursion, thus a positive excursion in the flux results in 
two negative excursions in the whitened flux, which are often misconstrued as transits
by the subsequent search. 

The removal of positive outliers is accomplished by marking their locations in the quarter-stitched flux
time series as gaps and applying the standard TPS gap filling algorithm. The identification of positive
outliers, by contrast, makes use of the whitened flux.  The advantage to this is that by design the whitened
flux contains only Gaussian-distributed, zero mean, unit variance white noise, plus quasi-impulsive outliers;
consequently, the positive outliers are extremely easy to identify in the whitened flux. The disadvantage is that
a positive outlier in the whitened flux can either indicate a positive outlier in the original flux, or it can be part
of the ring-down of a negative outlier such as a transit; this can be visualized by inverting the lower plot in 
Figure \ref{f2}. Thus the algorithm for positive outlier removal is as follows:
\begin{itemize}
\item whiten the quarter-stitched flux
\item identify clusters of whitened flux values which exceed a threshold: in this case a threshold of 12.3 $\sigma$ is
   used, as explained in Appendix \ref{outlier-threshold}
\item determine whether each cluster is due to positive outliers in the original flux or due to the ring-down of 
negative outliers in the original flux: this is accomplished by examining the local minima adjacent to each cluster, since
for positive outliers the local minima will be weaker than the positive outliers, whereas for the ring-down of a transit one 
of the local minima will be much stronger than the positive outliers
\item for each positive outlier value thus identified, mark the cadences in the quarter-stitched flux as gapped and
apply gap-filling
\item produce a new whitened flux from the outlier-removed quarter-stitched flux and iterate the process until no further
positive outliers are identified: this takes account of the fact that removal of outliers can change the local noise characteristics
slightly, causing values which had previously been below threshold to exceed the threshold.
\end{itemize}

\subsection{Limitation on Allowable Transit Duty Cycles}


An additional means of separating likely transiting planet signatures from false alarms is to apply bounds to the ratio 
of the transit duration $\tau$ to the orbital period $T$. The relationship between these parameters of the transit can
be derived from Kepler's laws of motion:
\begin{equation}\label{tau-T}
\tau = k T^{1/3}.
\end{equation}
We can rewrite Equation \ref{tau-T} in terms of the transit duty cycle $\phi_{\rm dut} \equiv \tau/T$:
\begin{equation}\label{phi-T}
\phi_{\rm dut} = k T^{-2/3}.
\end{equation}
The value of $k$ for a specific system is a function of the star's properties, and also the eccentricity of the orbit which
is under consideration. For circular orbits about the Sun, $k$ is approximately 0.058 days$^{2/3}$ \citep{GIL00}; for
a circular orbit about a late-type M dwarf star, $k$ is approximately 0.026 days$^{2/3}$. 

The shortest orbital period included in TPS searches is set to 0.5 days. At this limit, Equation \ref{phi-T}
shows that $\phi_{\rm dut}$ for a Sun-like star and a circular orbit is
approximately 0.092. To allow margin for elliptical orbits or stars which are far from Solar in their parameters, we limit
the maximum allowed value of $\phi_{\rm dut}$ to 0.16. This restriction is implemented by adjusting the minimum search
period for each trial transit duration used in the search: for 1.5-hour transits, the search is allowed to operate down to periods
of 0.5 days, while for 15-hour transits the minimum search period is limited to 3.9 days. This restriction was also enforced in 
the Q1-Q12 processing reported in \citet{pt2013}.

In the Q1-Q16 processing, an additional restriction was placed on the lower bound of allowed $\phi_{\rm dut}$ values, 
specifically
\begin{equation}\label{min-phi}
\phi_{\rm dut} \ge 0.017 T[{\rm days}]^{-2/3}.
\end{equation}
This limit is 3.4 times smaller than that expected for Solar stars and 1.5 times smaller than expected for late M dwarf stars, 
which allows margin for elliptical orbits, large impact
parameter values, and non-Solar parameters. Equation \ref{min-phi} sets a maximum search period which is a function
of transit duration: for example, 1.5-hour transits are limited to search periods of 50 days or less, while 3.0-hour transits are
limited to search periods of 300 days or less.

Note that the use of the duty cycle cuts creates an implicit trade-off between the purity of the search (i.e., rejection of
false positives) and its efficiency (i.e., acceptance of true positives). Specifically, the cuts will reject false positive 
detections around Sun-like stars on the basis of unphysicality. At the same time, the upper limit on permitted duty cycle
will reject some true positive detections on stars which are significantly larger than the Sun; 
the lower limit on permitted duty cycle will reject some true positive detections on stars which are significantly smaller than the
Sun; and both cuts can potentially reject true positive detections on Sun-like stars from planets with highly eccentric orbits.
It is our judgement that the regions of parameter space excluded by these cuts are acceptably small when both sides of
the trade-off are considered.

\subsection{Detection and Vetoing of Potential Signals}\label{vetoes}


Out of the 197,322 targets which were searched by TPS, a total of 112,981 were found to have at least
one periodic signal which exceeded the multiple event statistic of 7.1 $\sigma$, while 84,341 had no such
signal.  In this regard, current experience is consistent with past TPS analyses, in which the number of
targets for which the maximum multiple event statistic exceeded the detection threshold was unphysically
large, implying that the vast majority of these events are false alarms.  As in the past, a series of vetoes are
used to eliminate false alarms to the extent possible without rejecting excessive numbers of true transit signals.
The vetoes which are used in the current analysis -- a robust statistic and a series of vetoes based upon
$\chi^2$ statistics -- are described in modest  detail in \citet{pt2013}, and in particular the $\chi^2$ vetoes
are described in considerable detail in \citet{seader2012}, so only a description of changes to these quantities will 
be given here.

The multiple event statistic as well as the robust statistic only admit detections with three or more transits.  In 
detections where there are only three transits, the multiple event statistic is blind to the quality of each transit whereas 
the robust statistic scrutinizes each of the three transits to veto situations where one or more of the three transits 
overlaps significantly with a region of data that is anomalous in some way.  The algorithm for examining the transits for
this case has changed slightly from that employed to produce the Q1-Q12 results in \citet{pt2013}.  The past algorithm 
required that no more than 50\% of in-transit cadences be marked as anomalous for any of the three transits, whereas 
the current algorithm requires that the average of the in-transit data quality weights be at least 0.5.  Making this slight 
change has exposed an undesired sensitivity in the robust statistic algorithm.  This change is responsible for an
increase in long period false alarms which is discussed in Section \ref{detected-signals}.  Future development work will 
be directed at enabling the $\chi^2$ vetoes to identify these long period false alarms by including the data quality weights 
in the calculation of the degrees of 
freedom of the statistic (which are ultimately used in the computation of the reduced $\chi^2$).

The two versions of the $\chi^2$ vetoes described in \citet{pt2013} are again employed to produce the results of 
this paper.  The first of these, $\chi^2_{(1)}$, remains unchanged.  There are some subtle issues, however, associated with
the construction of $\chi^2_{(2)}$ which are discussed at length in \citet{seader2012} but will be briefly mentioned here
for completeness.  Correcting for these subtleties enhances the vetoing efficiency of $\chi^2_{(2)}$ by enabling both 
the quantities it is computed from and the statistic itself to have the correct statistical properties as described in 
both \citet{Allen:2004gu} and \citet{seader2012}.  
The first subtlety is that the $\chi^2_{(2)}$ calculation implicitly requires that the calculated noise properties of the flux time series are
not changed by the presence or absence of a transit signal. In fact, while the noise calculation is relatively robust against the
effect of transits, it is not formally invulnerable: while the presence or absence of a transit results in a change in the noise
estimate which is small enough to be neglected in the Multiple Event Statistic calculation, it was found to have an effect
on the value of $\chi^2_{(2)}$. This is addressed by applying the TPS auto-regressive gap-fill algorithm to the cadences 
which are in-transit, which produces a flux time series which is effectively transit-free; this time series is used to calculate
the noise properties of the flux time series for the purpose of $\chi^2$ discriminator calculations. The remaining subtleties
arise from the assumption that each transit is localized in time and isolated from every other transit (i.e., the transit model
consists of a series of short intervals of negative values separated by longer intervals of zero values). While this is true for the
unwhitened transit model, it is not true for the whitened transit model, which is the model which must be used in the calculation
of the discriminators: in the whitened domain, the transits are ``smeared,'' such that there are no cadences which have a model
value of zero, and thus the value of the model at one transit depends to greater or lesser extent on the values of all the other 
transits. To mitigate thse effects, the $\chi^2$ contribution from each transit is calculated with all other transits replaced by
the auto-regressive gap fill values used for the noise calculation, thus effectively performing the calculation for each transit
as though it was the only transit in the flux time series; in addition, the calculation neglects any cadence which is outside of
that transit, as determined by the unwhitened model, so that the effect of ``smearing'' into out-of-transit 
 cadence times is eliminated.

In addition to the above changes to existing vetoes, another version of the $\chi^2$ veto was implemented that is more 
akin to a classical $\chi^2$ and is described in great detail in \citet{seader2012} as $\chi^2_{(3)}$.  In this version, 
each single event statistic is compared with a calculated expectation value in the whitened time domain (to avoid the 
subleties mentioned above).  These differences are then summed according to the exact expression for a classical 
$\chi^2$, including division by the expectation value.  The thresholding on this statistic is done in the same manner as 
the previous two $\chi^2$ statistics as described in both \citet{pt2013} and \citet{seader2012}.  It was discovered 
through the course of analysis of these results that $\chi^2_{(3)}$ works well to veto short period false alarms but
the threshold used was too aggressive and is largely responsible for the cases of known short-period transits 
which were not detected in the most recent processing.  Work is currently 
underway to tune all the vetoes more appropriately. 

As mentioned above, the number of targets with a multiple event statistic above the detection threshold was
112,981.  The vetoes were then applied to this set of targets.
The robust statistic, with a threshold of $6.4 \sigma$, vetoed 64,233 target stars, leaving 48,748 targets for which there
was at least one signal which passed both the robust statistic and multiple statistic criteria.  The final layer
of vetoes based on $\chi^2$ statistics, all with thresholds of $7.0 \sigma$, removed 38,977 targets from consideration, leaving 9,771 target stars
which produced at least one threshold crossing event (TCE).  

\subsection{Detection of Multiple Planet Systems}

For the 9,771 target stars which were found to contain a threshold crossing event, additional TPS searches
were used to identify target stars which host multiple planets.  The process is described in \citet{hw2010} and in 
\citet{pt2013}. The multiple planet search incorporates a configurable upper limit on the number of TCEs per target, 
which is currently set to 10. This limit is incorporated for two reasons. First, limitations on available computing resources
translate to limits on the number of searches which can be accommodated, and also on the number of post-TPS
tests which can be accommodated. Second, applying a limit to the number of TCEs per target prevents a failure mode
in which a flux time series is so pathological that the search process becomes ``stuck,'' returning an effectively infinite
number of nominally-independent detections which all have the same ephemeris. The selected limit of 10 TCEs is
based on experience: to date, the maximum number of KOIs on a single target star is 7, which indicates that at this time
limiting the process to 10 TCEs per target is not sacrificing any potential KOIs.

The additional searches performed for detection of multiple planet systems yielded 6,573 additional TCEs across 3,229 target stars, for a
grand total of 16,344 TCEs. Note that all of these TCEs are subjected to the full TPS process of detection and
vetoing described above.

In the analyses below, a small number of the 16,344 TCEs are not included.  This is due to the desire to limit
the analysis presented here to TCEs for which there is a full analysis available from the Data Validation (DV) pipeline
module \citep{hw2010}.  A total of 28 targets, containing 55 TCEs, failed to complete their DV analyses and are thus
excluded.  A total of 4 targets produced 11 TCEs each, in excess of the 10 planet limit set as a user-specified parameter
for DV; in these cases, the eleventh TCE was reported but not included in the subsequent DV analysis.  Considering
these exclusions, a total of 9,743 targets produced 16,285 TCEs which are analyzed below. Only the 16,285 TCEs included
in this analysis will be exported to the tables maintained by the NASA Exoplanet 
Archive\footnote{http://exoplanetarchive.ipac.caltech.edu.}.

\section{Detected Signals of Potential Transiting Planets}\label{detected-signals}

\begin{figure}
\plotone{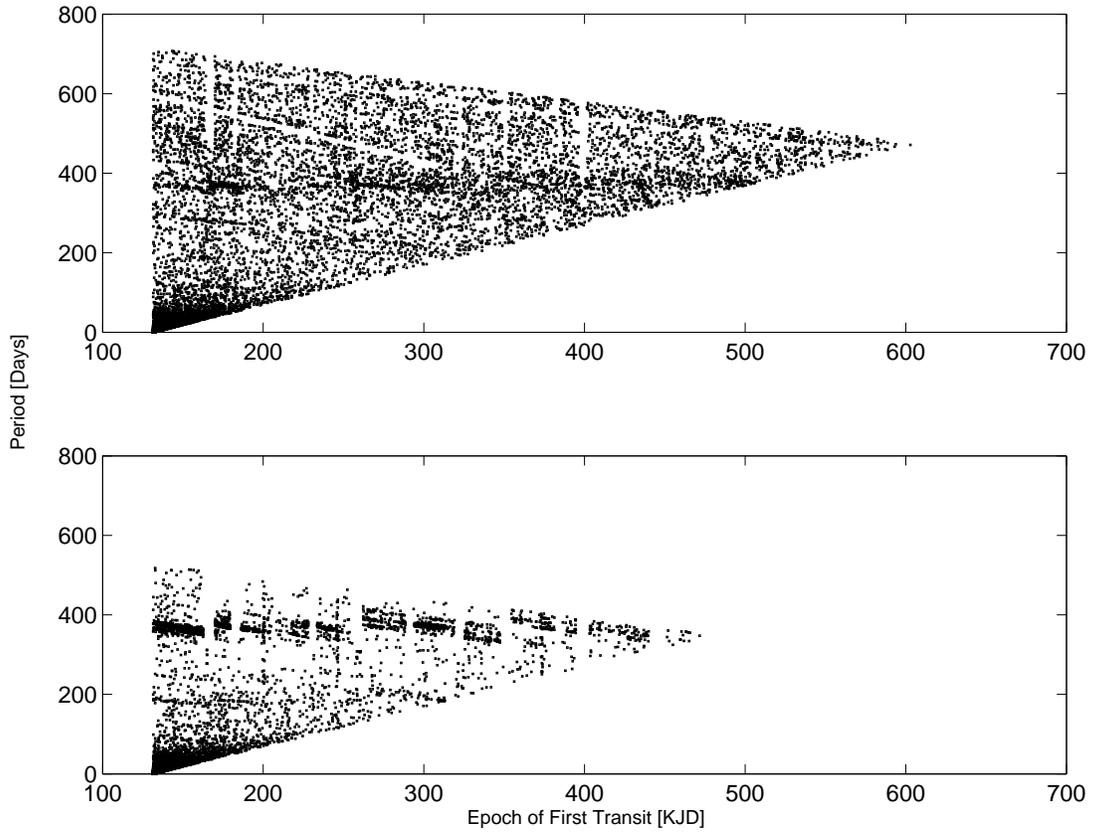}
\caption{Top: epoch and period of the 16,285 TCEs detected in Q1-Q16 of \kepler{} data; bottom: epoch and period of the
18,427 TCEs detected in Q1-Q12 of \kepler{} data, as reported in \citet{pt2013}.
Periods are in days, epochs are in
Kepler-modified Julian Date (KJD), see text for definition.
\label{f3}}
\end{figure}

As described above, a total of 9,743 targets in the \kepler{} dataset produced TCEs.  
For 6,520 of these targets, only one TCE was detected; for 3,223 targets, the multiple planet search detected additional
TCEs. The total number of TCEs detected across all targets was 16,285.
Figure \ref{f3} shows the period and epoch of each of the 16,285 TCEs, with period in days and epoch
in Kepler-Modified Julian Date (KJD), which is Julian Date - 2454833.0, the latter offset corresponding to January 1, 2009,
which was the year of \kepler{}'s launch. Figure \ref{f3} also shows the same plot for the 18,427 TCEs detected in the 12 quarter \kepler{}
dataset, as reported in \citet{pt2013}.  The axis scaling is identical for the two subplots, as is the marker size.  Several features
are apparent in this comparison.  First, the number of TCEs is reduced despite the fact that the number of targets and 
number of quarters of data have both increased since the earlier report, which demonstrates the improved false alarm vetoing
logic in the more recent analysis.  Second, as expected, the addition of 376 days of data acquisition has
increased the parameter space available for detections, as shown by the upward and rightward expansion of the ``wedge'' in 
Figure \ref{f3} from the Q1-Q12 to the Q1-Q16 results.
Third, the distribution of TCE periods appears to be more uniform for 
long periods in the current analysis, while in the previous one the number of TCEs decreased visibly at longer periods.

\begin{figure}
\plotone{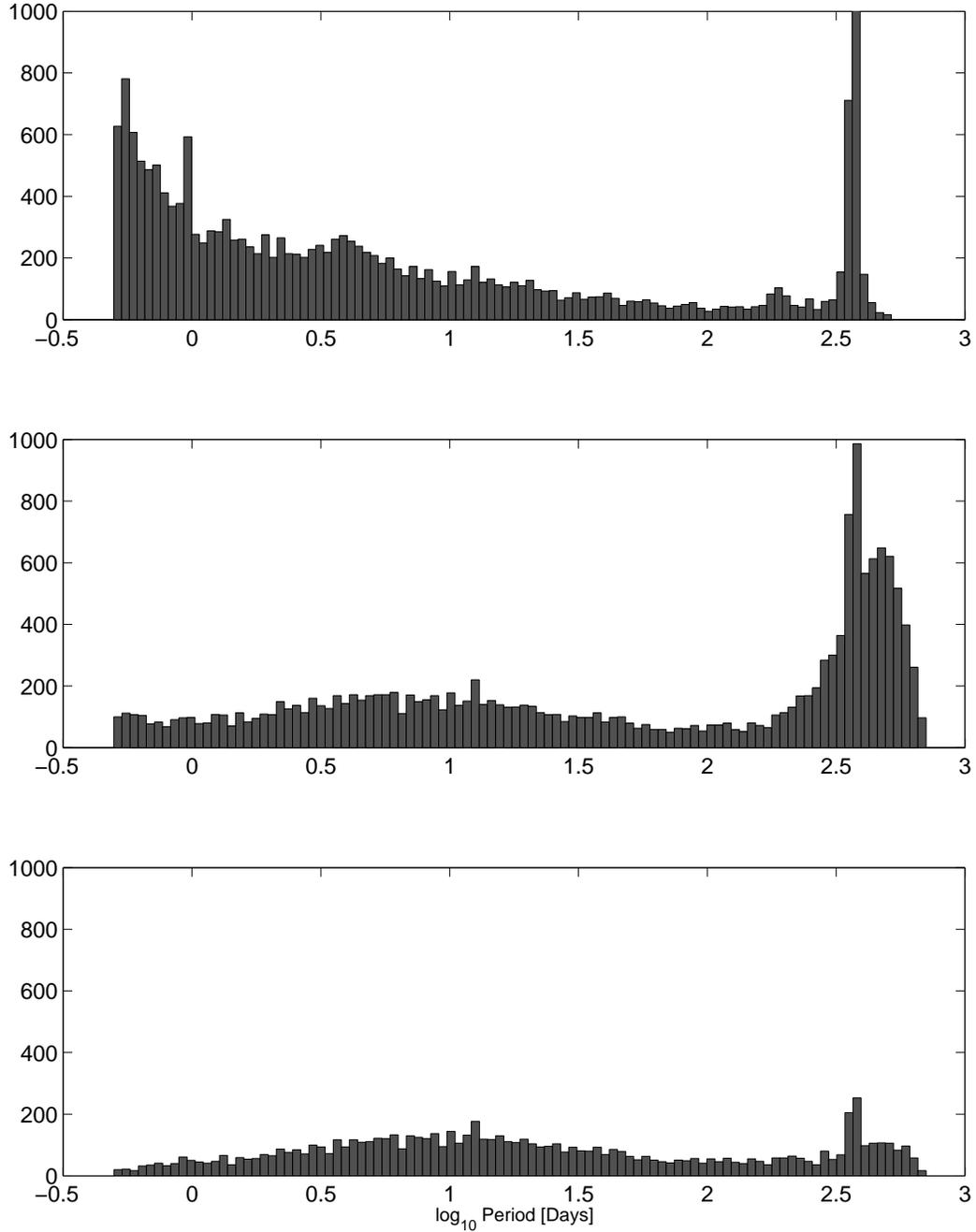}
\caption{Distribution of TCE periods, plotted logarithmically.  Top:  18,427 TCEs in the
Q1-Q12 search. Middle: 16,285 TCEs in the Q1-Q16 search. Bottom: 7,959 TCEs in the Q1-Q16 search which also
pass the additional criteria described in Section \ref{long-period-suppression}.
\label{f4}}
\end{figure}

The drastic change in the distribution of TCE periods can be seen more clearly in Figure \ref{f4}, which shows the distribution
of TCE periods on a logarithmic scale, with the Q1-Q12 results shown in the upper panel of the figure and the Q1-Q16
results in the middle panel.  The more recent search sharply reduces the number of short-period detections, which had
dominated the Q1-Q12 processing, but is in turn dominated by long-period detections. The increase in the number of 
long-period detections in Q1-Q16 compared to Q1-Q12 is vastly larger than would be expected from the addition of 4 
quarters of data, and appears to be due entirely to detections which contain 3 valid transits (i.e., there has been no corresponding
increase in the number of detections with 4, 5, 6, etc. transits, which would be expected if the new detections were all real
events which were made detectable by the additional data). Since the 3 transit events were the only ones affected by the 
change in the robust statistic described in Section \ref{vetoes}, our strong suspicion is that the large number of long-period
detections is dominated by false alarms which were inadvertently permitted to slip through the system by the aforementioned
change.

\begin{figure}
\plotone{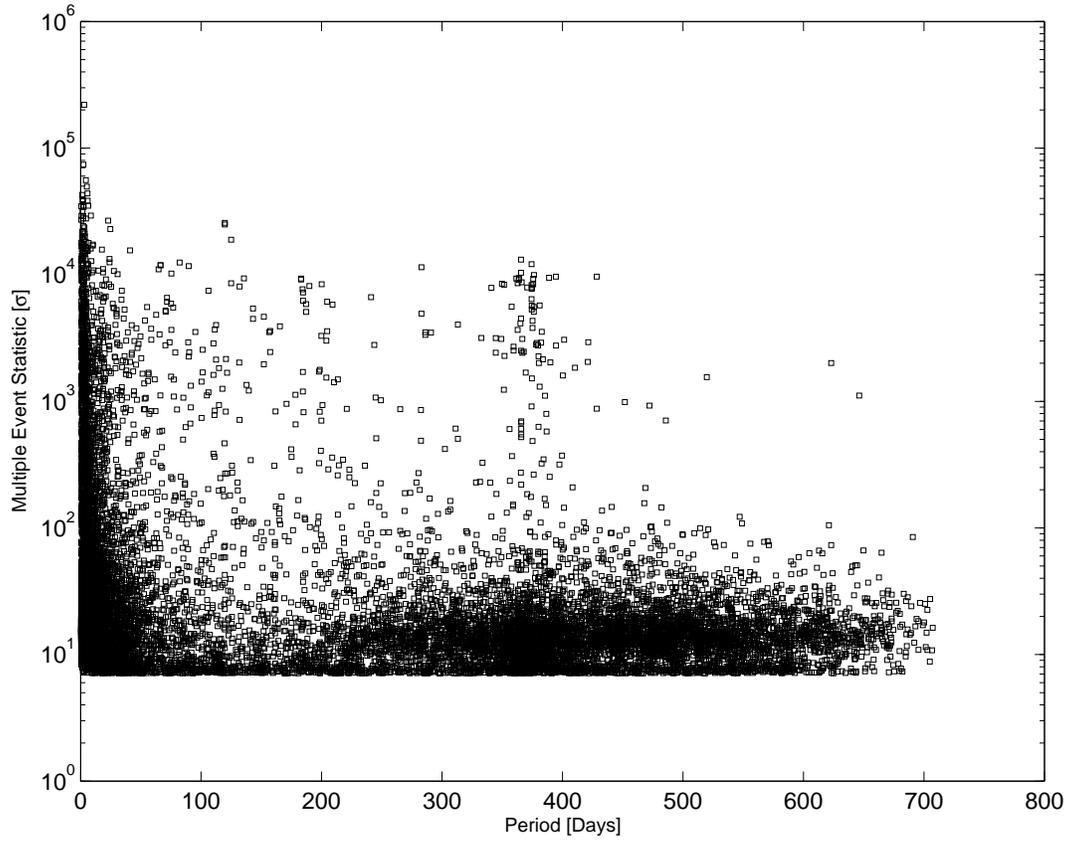}
\caption{Multiple event statistic and orbital period of 16,285 TCEs.
\label{f5}}
\end{figure}

\begin{figure}
\plotone{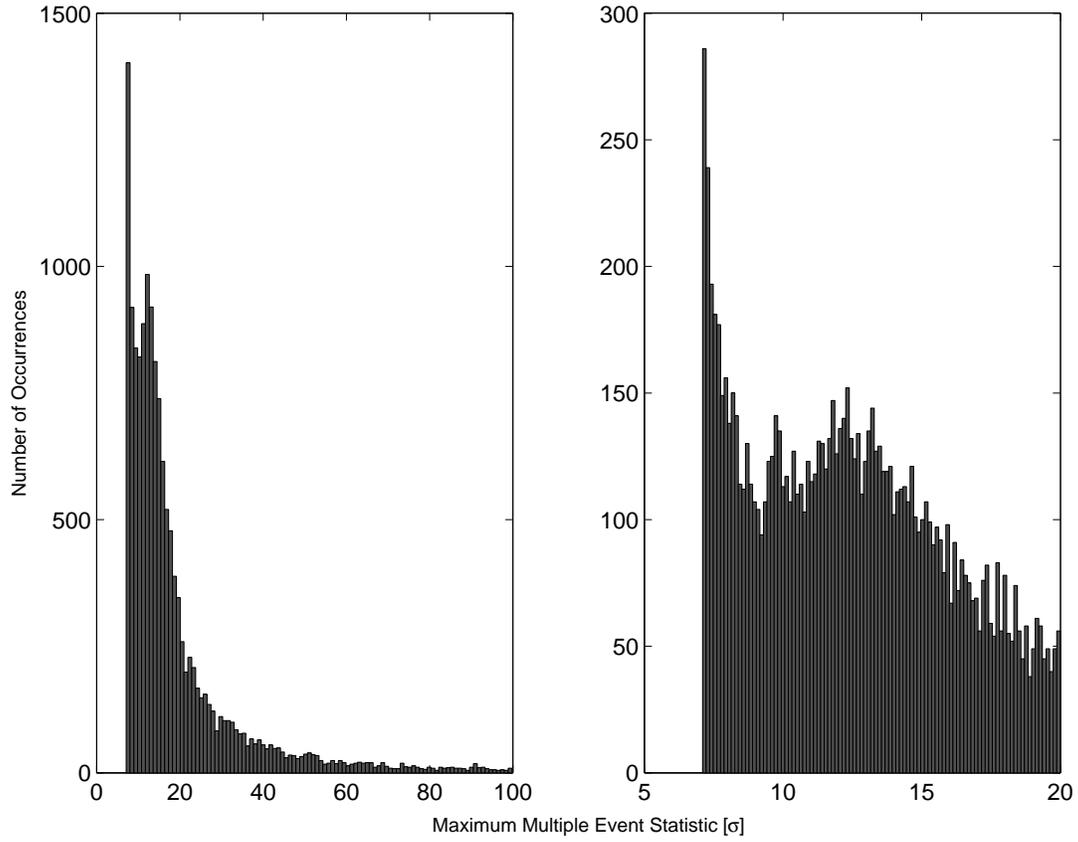}
\caption{Distribution of multiple event statistics.  Left:  14,506 TCEs with multiple event statistic below 100 $\sigma$.  Right:
10,651 TCEs with multiple event statistic below 20 $\sigma$.
\label{f6}}
\end{figure}

\begin{figure}
\plotone{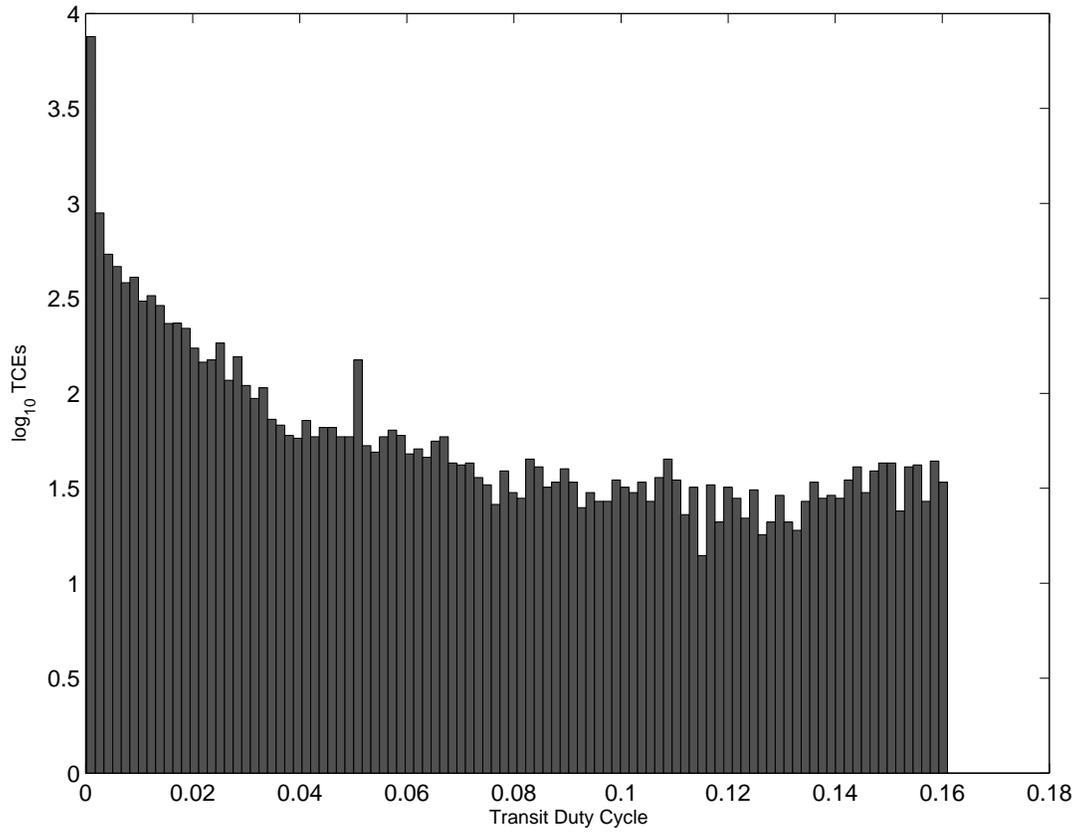}
\caption{Transit duty cycles of TCEs.
\label{f7}}
\end{figure}

Figure \ref{f5} shows the multiple event statistic (MES) and period of the 16,285 TCEs. The cluster of events with periods above
200 days, with relatively low multiple event statistic, are believed to be another representation of the long-period false alarms
discussed above. The relatively narrow cluster at approximately 380 days is due to the ``one \kepler{} year'' instrument artifact
discussed in \citet{pt2013}; this cluster is also visible in Figure \ref{f4}.
Figure \ref{f6} shows the distribution of
multiple event statistics: 14,506 TCEs with multiple event statistic below 100 $\sigma$ are represented in the left figure, while the
right hand figure shows the 10,651 TCEs with multiple event statistic below 20 $\sigma$.  Figure \ref{f7} shows the transit duty
cycles of the TCEs, with the vertical axis plotted on a logarithmic scale.  The strong overabundance of TCEs with extremely low
duty cycles is a consequence of the strong overabundance of TCEs with extremely long periods, as discussed above; by contrast
with the Q1-Q12 TCEs reported in \citet{pt2013}, the distribution is relatively flat from transit duty cycles of 0.04 to 0.16.  

\subsection{Suppression of Long Period False Alarm Detections}\label{long-period-suppression}

While the large number of likely false alarms at long period are not a problem in principle, in practice they represent a 
complication for the human-labor-intensive process of developing and classifying new 
\kepler{} Objects of Interest (KOIs) \citep{koi}.  Careful analysis
of the statistical properties of the TCEs, and in particular examination of the properties of re-detected known planet candidates
and comparison with the ensemble of new TCEs, has allowed the 
development of a series of ``back end'' discriminators which reject long-period TCEs that are likely to
be false alarms, while preserving long-period TCEs that are more likely to be astrophysical phenomena.

The first discriminator is the ratio of the multiple event statistic (MES) of a given TCE with the variance of the ensemble 
of MES values on the
same target, at the TCE period, but across all phases (MES\_MAD, so named because a median absolute deviation
is used to obtain a robust estimate of the variance).  
In the absence of any transits, the latter ensemble has an expected
mean value of zero and unit variance; the presence of transits and/or other low-duty-cycle phenomena results in a long, flat tail to the
MES distribution; the presence of more widespread systematic noise in the flux time series will increase the
variance of the MES ensemble.  Thus, rejecting cases in which the ratio of MES to MES\_MAD is low will remove cases in which the
flux time series noise properties are such that the TCE is suspect.

The second discriminator is a comparison of the MES and the signal-to-noise ratio (SNR) of the transit fit performed in
Data Validation.  The latter is expected to be higher than the former:  the Data Validation fit uses a properly-shaped limb-darkened transit model,
including fine adjustment of the transit duration, epoch, and period; the MES in the TCE is effectively the SNR achieved
by fitting a much lower-fidelity, box-shaped model to the same data.  When the ratio of the DV fit SNR to the TCE MES is low,
it indicates an event which does not have a transit-like shape. Note that, for some TCEs, no SNR value was available due to
issues in the Data Validation processing. In these cases, the cut on the SNR-to-MES ratio was not applied.

Finally, the MES of the TCE can be compared to the minimum MES on the same target at the TCE
period but across all epochs.  In the absence of astrophysical signatures, the MES follows a normal distribution with 
zero mean and unit variance; consequently, 
the minimum multiple event statistic at the TCE period should be a negative value, and should 
have a probability distribution given by the negative-valued portion of a Gaussian.
The ratio of the absolute value of the minimum multiple
event statistic (henceforth MES\_MIN) to the MES of the TCE
should therefore be small for a high-quality detection.  The presence of periodic positive excursions in the target flux time series at the TCE
period indicate a strong probability that some phenomenon other than transits is responsible for the TCE.

After analysis of the properties of the discriminators above, we found that it was possible to remove a
significant fraction of the long-period false alarms while preserving the high-quality TCEs. This is accomplished by rejecting
a TCE for which any of the following is true:

\begin{enumerate}
 \item MES / MES\_MAD $<$ 7.1,
 \item SNR / MES $<$ 0.6 (for TCEs which have an SNR value available), 
 \item MES\_MIN / MES $>$ 0.6.
\end{enumerate}

When these cuts were applied to the Q1-Q12 population of planet candidates, 
it was determined that approximately 1\% would be rejected and approximately 99\% retained.  Applying these cuts 
to the Q1-Q16 TCEs reduces the number of 
TCEs to be vetted from 16,285 to 7,959, and of particular importance in this run, from 6,073 to 1,243 with periods of 300 days
or more. The bottom panel in Figure \ref{f4} shows the distribution of periods for the 7,959 TCEs which passed the cuts listed above.

\subsection{Comparison with Known Kepler Objects of Interest (KOIs)}\label{koi-comparison}

As in past analyses \citep{pt2012,pt2013}, we have identified a subset of the Kepler Objects of Interest (KOIs) which we use as a
set of test subjects for the TPS run.  TPS does not receive any prior knowledge about detections on targets; therefore, the re-detection
of objects of interest which were previously detected and classified as valid planet candidates
 is a valuable test to guard against inadvertent introduction of significant flaws
into the detection algorithm.

The list of Q1-Q12 KOIs has been analyzed and a set of high-quality ``golden KOIs'' identified for comparison to the Q1-Q16 TCEs.  This subset
of the full KOI list is a representative cross-section of all KOIs in the parameters of transit depth, signal-to-noise, and period, and includes
cases which have been identified as eclipsing binaries or astrophysical false positives.

\begin{figure}
\plotone{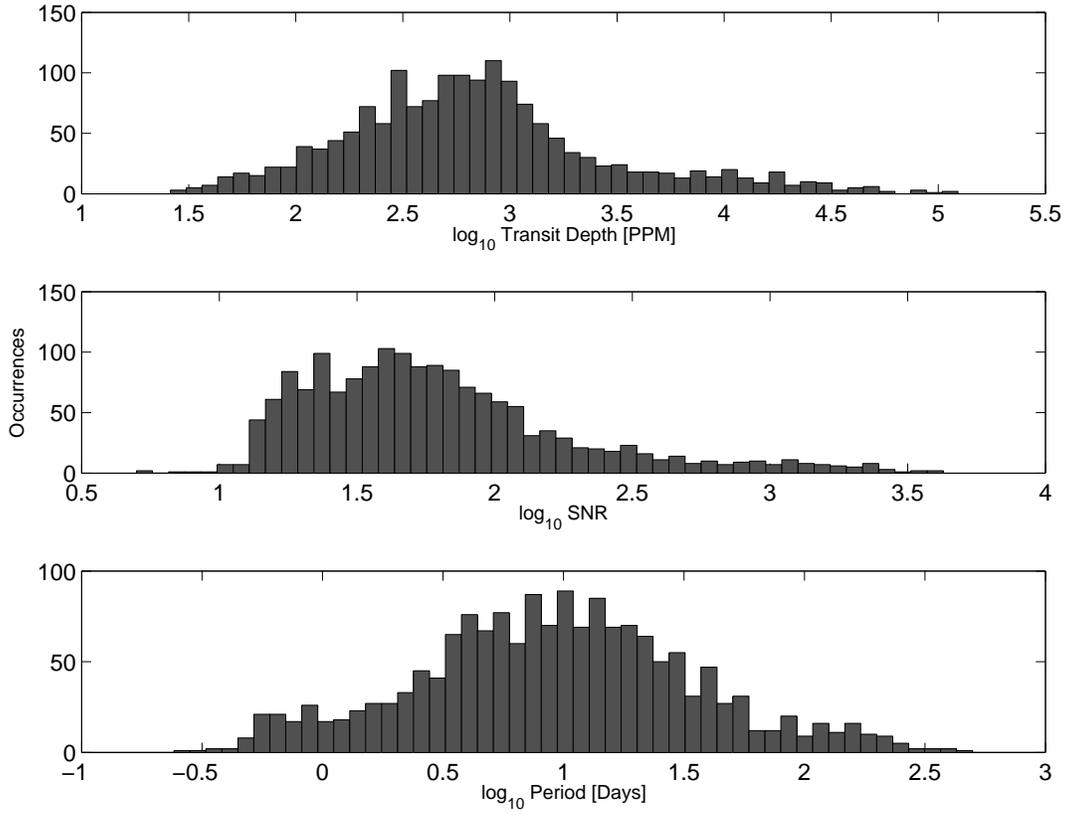}
\caption{Parameter distribution of ``golden KOIs.''  Note use of logarithmic horizontal axes in all cases.
\label{f8}}
\end{figure}

The ``golden KOI'' set includes 1,646 KOIs across 1,417 target stars.  Figure \ref{f8} shows the distribution of estimated transit depth,
signal-to-noise ratio, and period for the ``golden KOIs.''  Out of these, 1,372 target stars produced one or more TCE, while 45 target stars
did not.  All 45 of the target stars which produced no TCE have one and only one KOI per target, and the missed KOIs are strongly dominated
by short periods:  39 out of 45 have periods under 3 days, and only 1 out of 45 has a period in excess of 1 quarter.  Examination of the
short period failures indicates that they are dominated by a common failure mode, in which short period transits are mistaken for narrow-band
oscillations of the host star and eliminated by an algorithm which is designed to address such narrow-band oscillations (this algorithm is
discussed briefly in Section 2 of \citet{pt2012}). Note that the oscillations are only removed in instances in which they are strong, i.e., instances
in which a small number of narrow-band resonances dominate the stellar variability relative to more broad-band variations. As a consequence, the
removed transiting planet signatures have short periods and are strong relative to the background stellar variability. This implies that the 
resonance removal is primarily a problem for re-detection of planet candidate signatures which were detected early in the \kepler{} Mission, 
and has little relevance for detection of weak signals with short periods.

\subsubsection{Matching of KOI and TCE Ephemerides}

Detection of a TCE on a ``golden KOI'' target star is a necessary but not sufficient condition to conclude that TPS is functioning properly.  An
additional step is that the TCEs must be consistent with the expected signatures of the KOIs.  This is assessed by comparing the ephemerides
of the KOIs and their TCEs, as described in \citet{pt2013}; the ephemeris-matching process also implicitly compares the numbers of KOIs and
TCEs on each target star, which exposes cases in which, on a given star, some but not all KOIs were detected.

Of the 1,601 KOIs on the 1,372 target stars which produced TCEs, it was possible to find matches for 1,597 of the KOIs.  The two KOIs which did
not produce TCEs were KOI 1101.01 (KIC 3245969), and KOI 351.04 (KIC 11442793).  KOI 1101.01 is a short-period candidate (2.84 days), and was
most likely removed by the narrow-band oscillation algorithm; the second candidate on this target, with a period of 11.4 days, was detected with a correct
ephemeris match.  KOI 351 is a multi-planet system with significant transit timing variations (TTVs) \citep{koi351}; since TPS requires highly periodic
signals to produce a valid detection, its performance on this system has always been poor.  Nonetheless, TPS did detect KOI 351.03, KOI 351.05,
and KOI 351.06 with correct ephemeris matches, though in the case of 351.05 the match is only approximate (match criterion value of 0.82,
indicating overlap between 82\% of the transits in the KOI and the TCE).  

\begin{figure}
\plotone{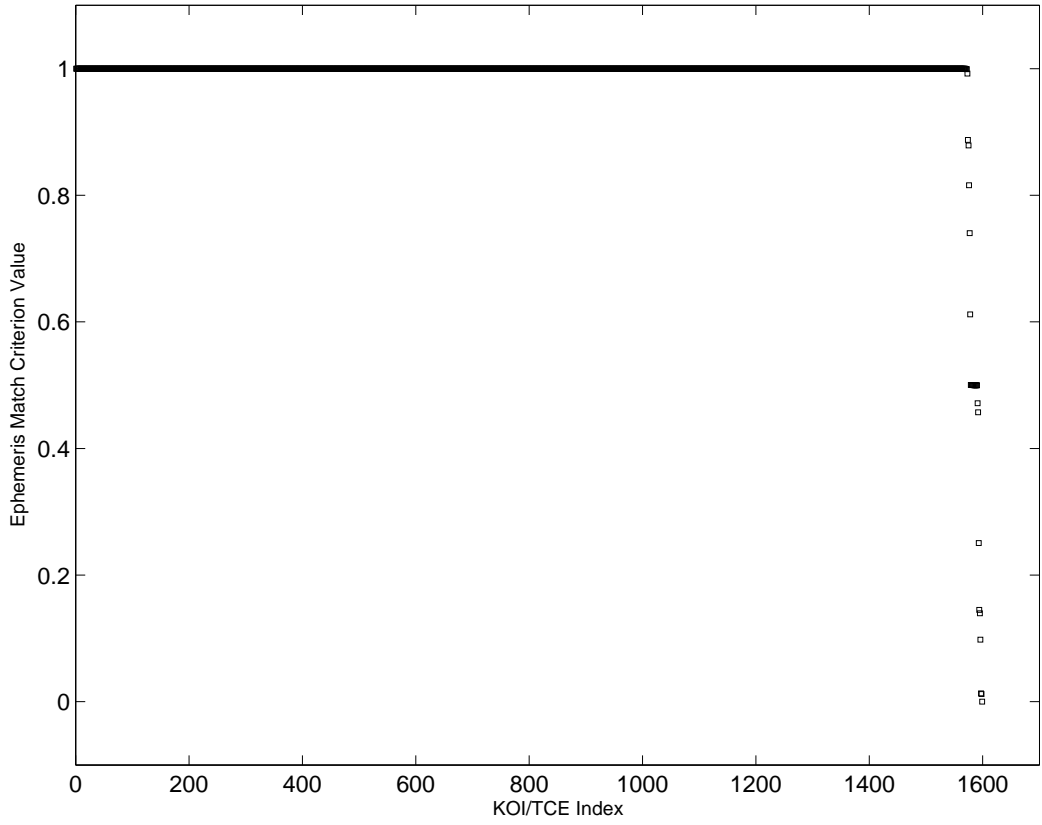}
\caption{Ephemeris match value for 1,599 KOI-TCE matches, sorted into descending order.
\label{f9}}
\end{figure}

Figure \ref{f9} shows the value of the ephemeris match criterion for the 1,597 KOI-TCE matches, sorted into descending order.  A total of 1,569 
KOI-TCE matches have a criterion value of 1.0, indicating that each transit predicted by one ephemeris corresponds to a transit predicted by the
other, to within one transit duration.  In these cases, it has been assumed that TPS correctly detected the ``golden KOI'' in question and no further
analysis was performed.

In the 28 cases in which the ephemeris match was imperfect, each KOI-TCE match was manually inspected.  The disposition of the results is as
follows:
\begin{itemize}
\item In 11 cases, the TCE actually matches the KOI, but the value of the match parameter does not reflect this; in general this is because the 
KOI ephemeris was derived with early flight data, requiring extrapolation to determine the transit times late in the mission and permitting an
accumulation of error in the KOI transit timings relative to the actual timings
\item In 13 cases, TPS detected a harmonic or sub-harmonic of the KOI, with a KOI period twice the TCE period the most common by far
\item In 2 cases, TPS detected the planet but produced an incorrect ephemeris due to transit timing variations (TTVs)
\item In 2 cases, TPS failed to detect the KOI and the TCE appears to be a false alarm.
\end{itemize}
In conclusion, out of the 28 KOI-TCE pairs which have imperfect ephemeris matches, only 2 actually constitute a failure of the detection algorithm.

In addition to the TCEs described above, there were 476 TCEs detected on the KOI targets which are not on the list of ``golden KOIs.''  The majority
of these are known KOIs which were not included on the ``golden KOI'' list (i.e., cases in which some of the KOIs on a given target star were included
while others fell below the threshold for inclusion); in other cases, the unmatched TCE is a secondary eclipse of an eclipsing binary or an occultation
of a large planet behind its host star.  In a few cases, these may constitute new detections of additional planet candidates on stars already known to 
host one or more such candidates.

\subsubsection{Conclusion of TCE-KOI Comparison}

Out of 1,646 ``golden KOIs'' used to demonstrate the validity of the TPS algorithm, 1595 were correctly detected, for a recovery rate of 96.9\%.  
The missed KOIs were largely overlooked by TPS due to an algorithm which, in removing from the data narrow-band resonances due to stellar variation,
occasionally removes short-period transiting planet signatures. This removal issue is not expected to impact future discoveries of transit signatures
which are weak relative to the overall stellar variability of their host stars.

\section{Conclusions}

The Transiting Planet Search (TPS) pipeline module was used to search photometry data for 197,320 \kepler{} targets acquired over  4
years of science operations.  This resulted in the detection of 16,285 threshold crossing events (TCEs) on 9,743 target stars.  The 
distribution of TCEs was qualitatively different from those
obtained in a similar search utilizing 3 years of data: the more recent analysis contains a larger proportion of long-period detections which are considered
likely false alarms, but a smaller proportion of short-period false alarms.  The differences are believed to be due to changes made to the TPS algorithm, 
rather than to the additional flight data or changes in the data pre-processing algorithms.  Out of 1,646 Kepler Objects of Interest (KOIs) used to test the
detection and veto algorithms, 1,595 were correctly detected; the missed detections were dominated by short-period objects, and a known limitation of TPS is
suspected in these cases.

\section{Acknowledgements}

Funding for this mission is provided by NASA's Space Mission Directorate.  The
contributions of Hema Chandrasekaran continue to be essential in the studies documented
here.
 \appendix

\section{Determining the Threshold for Positive Outlier Removal}\label{outlier-threshold}

In Section \ref{remove-positive-outliers}, an algorithm is described which removes positive outliers from each flux time series
prior to searching same for transits. The algorithm requires a threshold for removal of the outliers, which is set at 12.3 $\sigma$. 
The rationale behind this value is explained below.

The most obvious requirement for the positive outlier threshold is that it should, indeed, address only outliers and not flux values
which are merely above the mean value due to statistical fluctuations. For a whitened flux time series with 69,810 samples, in the 
limit of Gaussian statistics, we expect that, on average, 1 sample will lie 4.2 $\sigma$ above the mean. Therefore, any threshold which
is significantly higher than 4.2 $\sigma$ can beu expected to be harmless in terms of its effect on sample values which are driven 
by statistical fluctuations alone.

Another consideration is that the purpose of removing positive outliers is actually to suppress the formation of transit-like features
in the ring-down of the outliers, since the latter can lead to false alarm detections. Given a detection threshold of 7.1 $\sigma$ and
a requirement of at least 3 transits for detection, a negative feature with a significance of $7.1 \sigma \times \sqrt{3} = 12.3 \sigma$ can
exceed the detection threshold when paired with two ``transits'' which each have a significance of zero. This implies that a transit-like
feature with a significance of 12.3 $\sigma$ has a strong probability of triggering a false alarm, and thus our threshold for removing
positive outliers should be set such that any outlier which is likely to have a 12.3 $\sigma$ transit-like feature in its ring-down 
is removed.

Since the ring-down of a positive outlier will always be smaller than the outlier itself, it follows that removing positive outliers with
12.3 $\sigma$ significance is more than sufficient to prevent formation of negative ring-down features with 12.3 $\sigma$ significance.
Since 12.3 $\sigma$ is much larger than 4.2 $\sigma$, it also follows that this threshold will be benign from the point of view of 
ignoring statistical fluctuations. Thus a threshold of 12.3 $\sigma$ was adopted for positive outlier removal.
 

\clearpage

\clearpage

\clearpage

\clearpage

\clearpage

\clearpage

\clearpage

\clearpage

\clearpage

\end{document}